\documentstyle[psfig,twocolumn]{article}

\def\thetitle {CAM-8: a computer architecture based on cellular automata}

\title {\thetitle\thanks {This research was supported by the Advanced
Research Projects Agency, grant N0014-89-J-1988.}}

\author {Norman Margolus
	\\ MIT Laboratory For Computer Science
	\\ Cambridge Massachusetts 02139}

\date{December 15, 1993}

 \newcommand{\figgiven}[3]{% label, picture, caption
 \begin{figure}[htbp]
  $$#2$$\relax
  \caption{#3}
 \label{fig.#1}
 \end{figure}                 }

 \newcommand{\figtwogiven}[3]{% label, picture, caption
 \begin{figure*}[htbp]
  $$#2$$\relax
  \caption{#3}
 \label{fig.#1}
 \end{figure*}                 }

\def\asic{{\sc asic}}
\def\cam{{\sc cam}}
\def\Cam{{\sc Cam}}
\def\sram{{\sc sram}}

\def\dram{{\sc dram}}
\def\fpga{{\sc fpga}}

\begin{document}

\maketitle

\begin{abstract}

The maximum computational density allowed by the laws of physics is
available only in a format that mimics the basic spatial locality of
physical law.  Fine-grained uniform computations with this kind of
local interconnectivity (Cellular Automata) are particularly good
candidates for efficient and massive micro-physical implementation.

Conventional computers are ill suited to run CA models, and so
discourage their development.  Nevertheless, we have recently seen
examples of interesting physical systems for which the best
computational models are cellular automata running on ordinary
computers.  By simply rearranging the same quantity and quality of
hardware as one might find in a low-end workstation today, we have
made a low-cost CA multiprocessor that is about as good at large CA
calculations as any existing supercomputer.  This machine's
architecture is scalable in size (and performance) by orders of
magnitude, since its 3D spatial mesh organization is indefinitely
extendable.

Using a relatively small degree of parallelism, such machines make
possible a level of performance at CA calculations much superior to
that of existing supercomputers, but vastly inferior to what a fully
parallel CA machine could achieve.  By creating an intermediate
hardware platform that makes a broad range of new CA algorithms
practical for real applications, we hope to whet the appetite of
researchers for the astronomical computing power that can be harnessed
in microphysics in a CA format.
\end{abstract}

\maketitle

\section{Introduction}

Within the Information Mechanics Group at the MIT Laboratory for
Computer Science, a primary focus of our research has been on the
question: ``How can computations and computers best be adapted to the
constraints and opportunities afforded by microscopic physics?''  This
has led us to study spatially organized computations, since the
maximum computational density allowed by the laws of physics is
available only in a format that mimics the basic spatial locality of
physical law.  Fine-grained uniform computations with this kind of
local interconnectivity (Cellular Automata) are particularly good
candidates for efficient and massive micro-physical implementation.

We have been involved for over a decade in the design and use of
machines optimized for studying Cellular Automata (CA) computations
\cite{cam5,margolus-bbm,super,cambook,cam8-book}.  This
involvement began in response to our need for more powerful CA
simulation tools---suitable for investigating the large-scale behavior
of CA systems.  Using our early CA machines (\cam{}s) we developed a
number of new CA models and modeling techniques for physics and for
spatially-structured computation \cite{cambook}.  Eventually we
``published'' a commercial version of our CA machines, along with a
collection of models as software examples \cite{cambook,cam6-soft}.

Some of our earliest models were reversible {\em lattice gases} that
simulated a billiard-ball computer \cite{margolus-bbm}.  It was a
natural step to use these and related lattice gases to try to simulate
fluid flow \cite{pde}.  Although only the linear hydrodynamics worked
correctly (see Figure~\ref{fig.hpp-code}), our \cam{} simulations made
Pomeau and others realize that lattice gases were not just conceptual
models, but might be turned into powerful computational tools (cf.\
the seminal ``FHP'' lattice-gas paper \cite{fhp}, and our companion
paper \cite{super}).

The design of our latest CA machine, \cam-8, builds upon our
accumulated experience with previous cellular automata machine
designs, and represents both a conceptual and practical breakthrough
in our understanding of how to efficiently simulate CA systems
\cite{patent,cam8-book}.  This new machine is an indefinitely
scalable three-dimensional mesh-network multiprocessor optimized for
large inexpensive simulations, rather than for ultimate performance.
Our small-scale prototype---with an amount and kind of hardware
comparable to that in a low-end workstation---already performs a wide
range of CA simulations at speeds comparable to the best numbers
reported for any
supercomputer \cite{chen-fast,kohring,yaneer-poly,yepez-cip}.  Machines
orders of magnitude bigger and proportionately faster can be built
immediately.

Most of the current exploration of cellular automata as computational
models for science is being done using machines which were designed
for very different purposes.  Such experimentation doesn't make
apparent the tremendous computational power that is potentially
available to models tailored for uniform arrays of simple processors.
Nevertheless, we already have seen examples of interesting physical
systems for which the best computational models are cellular automata
running on ordinary computers
(cf. \cite{kadanoff89,rothman90a,rothman92}).  \Cam-8---using a
relatively small degree of parallelism---makes possible a level of
performance at CA calculations much superior to that of existing
supercomputers, but vastly inferior to what a fully parallel CA
machine could achieve.  By creating an intermediate hardware platform
that makes a broad range of new CA algorithms practical for real
applications, we hope to whet the appetite of researchers for the
astronomical computing power that can be harnessed in microphysics in
a CA format.

\section{An architecture based on cellular automata}

\figtwogiven{system}{ \hfill\psfig{figure=system.ps,height=2in} \hfill
}{\Cam-8 system diagram. (a) A single processing node, with \dram{}
site data flowing through an \sram{} lookup table and back into \dram.
(b) Spatial array of \cam-8 nodes, with nearest-neighbor (mesh)
interconnect (one wire per bit-slice in each direction).}

\noindent
In nature, we have a uniform and local law in the world that is
operating everywhere in parallel.  A CA model is a synchronous digital
analog of such a law.  As a basis for a computer architecture, CA's
have the advantage that there can be a direct mapping between the
computation and its physical implementation: a small region of the
computer can implement a small region of the CA space, and adjacent
regions of physical space can implement adjacent regions of the CA
space.  Thus locality is preserved, and very efficient realizations
are in principle possible.  This efficiency, however, comes at the
cost of requiring that all models run on the machine must be spatially
organized.  Thus the unavoidable problem of ultimately making your
computation fit into a uniform and local physical world is shifted
into the software domain: you must directly embed your software
problems into a uniform and local spatial matrix.

\Cam-8 is a parallel computer built on this spatial paradigm.  For
technological convenience, it time-shares individual processors over
``chunks'' of space---and also time-shares the wires connecting each
processor with its neighboring processors.  The time-sharing of
communication resources reduces the number of interprocessor wires
dramatically and thus allows the scalability that is inherent in the
CA paradigm to be practically achieved using current technology, even
in three dimensions.  The time-sharing of processors allows a highly
efficient ``assembly-line'' processing of spatial data, in which
exactly the same operations are repeated for every spatial site in a
predetermined order.

{}From the viewpoint of the programmer, this {\em virtualization} of
the spatial sites is not apparent: you simply program the local
dynamics in a uniform CA space.

\subsection{System overview}

Figure~\ref{fig.system} is a schematic diagram of a \cam-8 system.  On
the left is a single hardware module---the elementary ``chunk'' of the
architecture.  On the right is an indefinitely extendable array of
modules (drawn for convenience as two-dimensional, the array is
normally three-dimensional).  A uniform spatial calculation is divided
up evenly among these modules, with each module simulating a volume of
up to millions of fine-grained spatial sites in a sequential fashion.

In the diagram, the solid lines between modules indicate a local {\em
mesh} interconnection.  These wires are used for spatial data
movement.  There is also a tree network (not shown) connecting all
modules to the front-end workstation that controls the CA machine.
The workstation uses this tree to broadcast simulation parameters to
some or all modules, and to read back data from selected modules.
Normally, the parameters of the next updating scan of the space are
broadcast while the current scan is in progress, and analysis data
{}from the modules are also read back while the current scan runs.

Each module contains a separate copy of the current program for
updating the space (data transformation parameters, data movement
parameters, etc.), and all modules operate in lockstep.  This allows
both the computation within modules and communication between modules
to be pipelined, so that one virtual processor within each module
completes its update (including all communication) at each machine
clock.

Spatial site data is kept in conventional \dram{} chips which are all
accessed continuously in a predictable and optimized scan order,
achieving 100\% utilization of the available memory bandwidth.  Within
a module, each \dram{} chip belongs to a separate {\em bit-slice}, and
each \dram{} chip has its address controlled separately from the rest.
The group of bits that are scanned simultaneously (one bit from each
bit-slice) constitute a {\em hardware cell}.  Data is reshuffled
between hardware cells by controlling the relative scan order of the
\dram{} bit-slices.

Updating is by table lookup.  Data comes out of the cell-array,
is passed through a lookup table, and put back exactly where it came
{}from (Figure~\ref{fig.system}a).  The lookup tables are double
buffered, so that the front-end workstation can send a new table while
the \cam-modules are busy using the previous table to update the
space.  There are also hardware provisions for replacing the lookup
tables with pipelined logic (to allow versions of \cam-8 with a large
number of bits in the hardware cell---too many to update by table
lookup), and for connecting external data sources or analysis
hardware.

There are only a handful of connections between modules---one per
bit-slice to each of the six adjacent modules.  Uniform data shifts
across the entire three-dimensional space are achieved by combining
\dram{} address manipulation with static
routing \cite{margolus-thesis,static}: data are sent over the
intermodule wires at preordained times, exactly when they are needed
by adjacent modules.

\medskip
\subsubsection{A sample implementation}\label{technology}

For comparison purposes, here is a description of the amount and kind
of technology used in one of our prototype 8-module \cam-8 units:

\begin{itemize}
	\item {\bf System clock: } 25 MHz
	\item {\bf DRAM: } 64 Megabytes (4 Megabit\\ chips, 70ns)
	\item {\bf SRAM: } 2 Megabytes (256 Kilobit\\ chips, 20ns)
	\item {\bf Logic: } about 2 Million gates total
	\item {\bf Logic technology: } 1.2 micron CMOS
\end{itemize}

\noindent
This level of technology is comparable to what is used in a low-end
work\-station---a small \cam-8 unit is really a CA {\em personal
computer}.\footnote{In comparing the performance of this unit against
numbers reported for simulations on supercomputers (which have a
similar performance) one should also take availability into account: a
personal computer can be run on a single problem for a very long
period of time.  Economies of scale (mass production) are also
potentially available to personal-computer level hardware.} For CA
rules with one bit per site, this 8 module machine runs simulations at
a rate of about 3 billion site updates per second on spaces of up to
half a billion sites; with 16 bits per site, simulations run at about
200 million site updates per second on spaces of up to 32 million
sites.  Several of our 8-module prototypes can be connected together
to construct bigger machines---repackaging the modules would be
desirable for constructing substantially larger machines.

Our \cam-8 prototype can directly accumulate and format data for a
real-time video display; provision is also made to accept data
directly from a video camera, in order to allow \cam-8 to perform
real-time video processing with CA rules.  For a detailed description
of the prototype \cam-8 implementation, including the \cam-8 register
model, the workstation interface, and system configuration and
initialization, see ``STEP: a Space Time Event
Processor \cite{step1}.''

\subsection{Programmer's model}

In addition to more specialized resources having to do with display,
analysis, and I/O, the main programmable resources in \cam-8 are:

\begin{itemize}
	\item Number of dimensions.
	\item Size and shape of the space.
	\item Number of bits at a site.
	\item Initial state of the space.
	\item Directions and distances of data\\ movement.
	\item Rules for data interaction.
\end{itemize}

\noindent
All of these parameters are normally specified as part of a \cam-8
experiment.  Often the data movement and data interaction will change
with time, either cyclically or progressively as the simulation runs:
the overhead associated with changing these parameters before every
update of the space is negligible.

\medskip
\subsubsection{The space}

Our earlier \cam{} machines were all 2-dimensional, with severe
restrictions on the overall size of the space and the number of bits
at each spatial site.  In \cam-8, these parameters may be freely
specified.

The overall space-array is configured as a multi-dimensional Cartesian
lattice with a chosen size, shape, number of bits per site, and number
of dimensions.  The boundaries are periodic---if you move from site to
site along any dimension, you eventually get back to your starting
point.  Three of the dimensions can be arbitrarily extended by adding
``chunks'' of hardware ({\em modules}).  The maximum number of bits in
the array is of course governed by the total amount of storage in all
of the modules (64 Megabytes in our prototype): each module processes
an equal fraction of the overall space-array.  There is no
architectural limit on how many modules a \cam-8 machine can have.

\medskip
\subsubsection{Data movement}

In earlier \cam{} machines, there were severe constraints on {\em
neighborhoods}: restrictions on which data from sites near a given
site could be seen by the CA update rule acting at that site.  In
\cam-8, we have eliminated these constraints.  This was accomplished
by abandoning the use of traditional CA neighborhoods, and basing our
machine on the kind of data partitioning characteristic of lattice gas
models.  Instead of having a fixed set of neighborhood data visible
{}from each site, we shift data around in our space in order to
communicate information from one place to another.

In traditional CA rules, each bit at a given site is visible to all
neighbors.  In contrast, the pure data-movement used in \cam-8 sends
each bit in only one direction.  {\em Information fields} move
uniformly in various directions, each carrying corresponding bits from
every spatial site along with them---in two dimensions think of
uniformly shifting bit-planes, in higher dimensions
bit-hyperplanes.\footnote {The term {\em information field} is a bit
of a pun, since we intend by this both the computer science meaning,
namely a fixed set of bits in every ``record'' (spatial site), and
also the physics meaning of a field, which is a number attached to
each site in space.} Interactions act separately on the data that land
at each lattice point, transforming this block of data into a new set
of bits.  If some piece of data needs to be sent in two directions,
the interaction must make two copies.

There is a constraint on how far bit-fields can move in one updating
step, but it is quite mild.  Each bit-field can independently shift by
a large amount in any direction---the maximum shift-component along
each dimension is one that would transfer the entire {\em sector} of a
bit-field contained in one hardware module into an adjacent module.
For a two dimensional simulation on our prototype, for example, the
$x$ and $y$ offsets for each bit-field that can be incorporated as
part of a single updating step can be any pair of signed integers with
magnitudes of up to a few {\em thousand}.  In general (for any number
of dimensions), each updating event brings together a selection of
bits chosen from the few {\em million} neighboring sites.

\medskip
\subsubsection{Data interaction}

Data movement and data interaction alternate: once we have all the
data in the right place, we update each site using only the
information present at that site.\footnote{Actually, the hardware does
both movement and updating in a single pipelined operation.} In our
prototype, there is a constraint that only 16 bit-fields can be moved
in independent directions simultaneously, and only 16 bits at a time
can interact and be updated arbitrarily (by table lookup).\footnote
{Alternative implementations (using the same \cam{} data-movement
chips) would allow many more simultaneously moving bit-fields, but
would use pipelined logic in place of lookup tables, since tables grow
in size exponentially with the number of inputs.  Sufficiently wide
programmable logic can perform any desired many-input function if
there are enough levels of logic; an arbitrary number of levels can be
simulated by changing the program for the logic from one scan of the
space to the next.  An interesting application of this would be for
efficiently running lattice gases with large numbers of bits per site
(cf. \cite{fung}).}

Thus a program for this machine consists of a sequence of
specifications of (wide ranging) particle-like data movements and
(arbitrary) 16-bit interaction events.  Simulations involving the
interaction of large numbers of bits at each site have to be broken
down into a sequence of 16-bit events---a space-time event program.

\section{Applications}

\Cam-8 is good at spatially moving data, and at making the data
interact at lattice sites.  This makes it well suited for simulating
physical systems using lattice-gas-like dynamics.  This also makes it
appropriate for a wide range of other spatially organized calculations
involving localized interactions.

We are actively collaborating with several groups to develop sample
applications which illustrate the use of this CA machine for physical
simulations (e.g., fluid flow, chemical reactions, polymer dynamics),
two and three dimensional image processing (e.g., document reading,
medical imaging from 3D data), and large logic simulations (including
the simulation of highly parallel CA machines).  Of course all of the
models developed for our earlier \cam{} machines \cite{cambook} run
well on \cam-8, and can now be extended far beyond the capabilities of
these earlier machines.  Many spatial algorithms (systolic,
{\sc simd}, etc.) designed for other machines \cite{static,ftl} can
also be adapted to this architecture.

As illustrations of the use of \cam-8, some sample applications and
simulation techniques are discussed below.  All of these examples have
been developed on the prototype machine discussed in
Section~\ref{technology}, and performance figures are for this
workstation-scale device.

\subsection{Lattice gases}

\figtwogiven{fhp7}{ \hfill%
\vbox{%
\psfig{figure=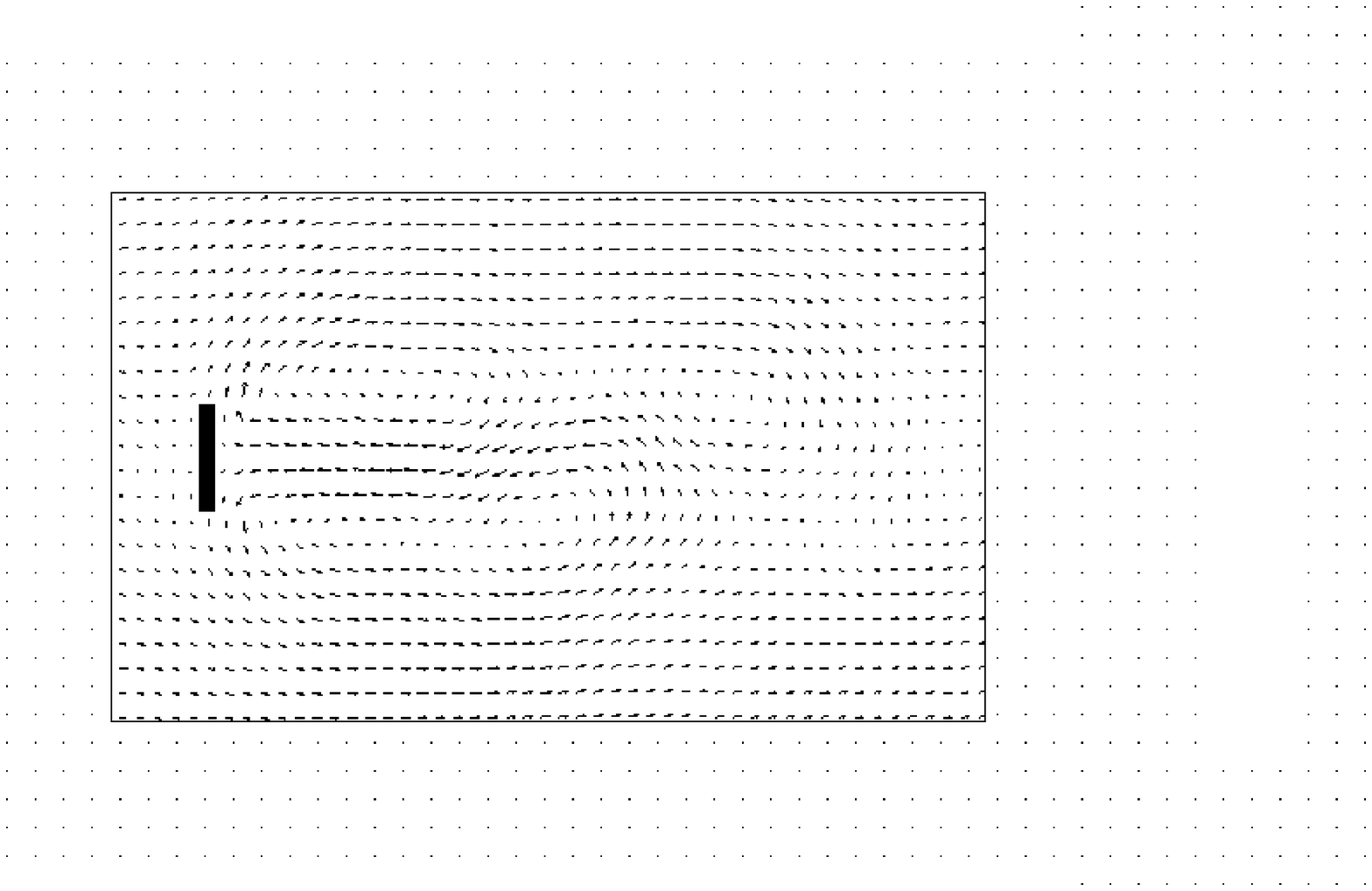,width=2.4in}%
}%
\hskip .16in%
\vbox{%
\psfig{figure=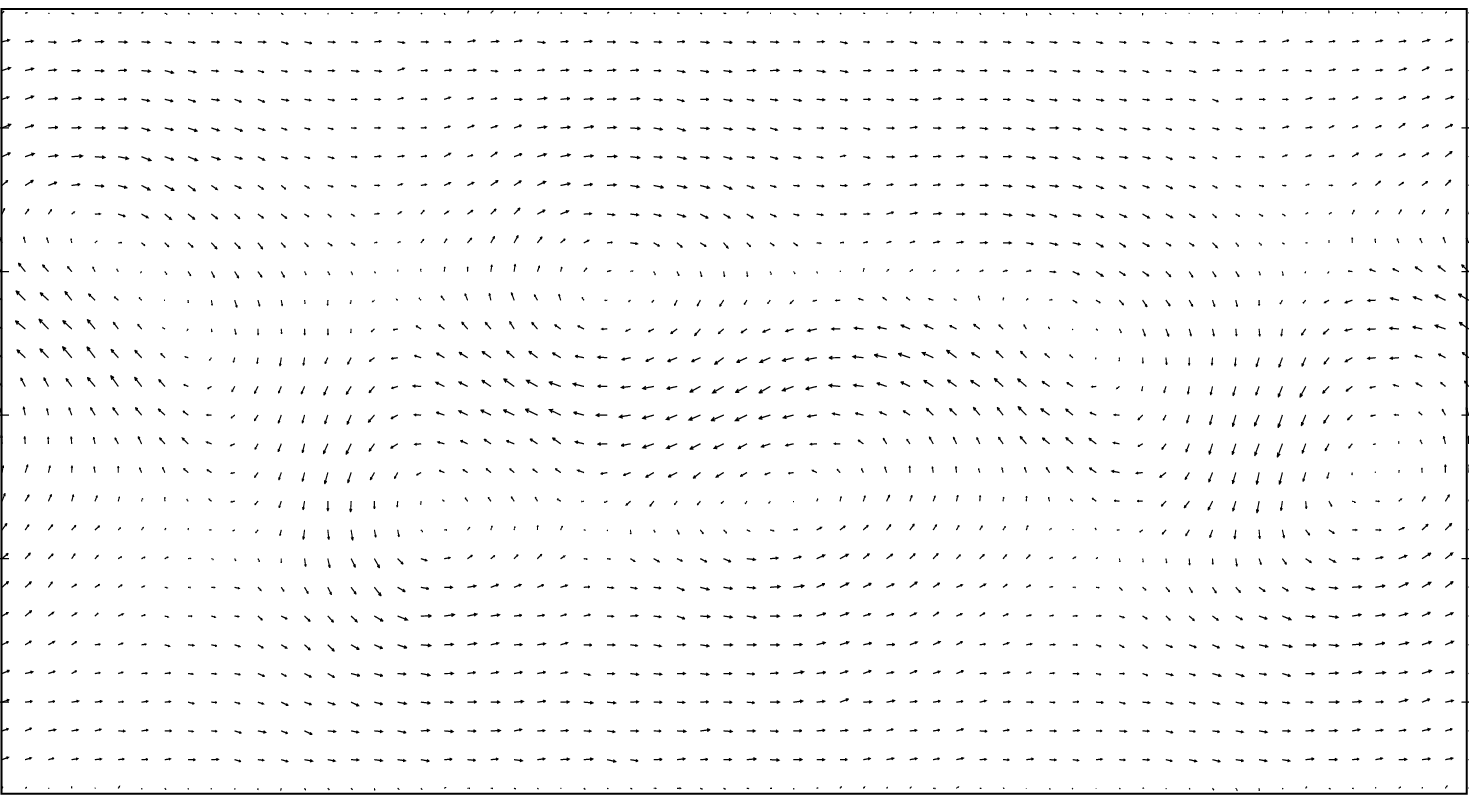,width=2.4in,height=1.437in}%
}%
\hfill }{Flows for two simulations using the FHP lattice gas.  (a)~Von
Karman streets,  (b)~Kelvin-Helmholtz shear instability.}

\noindent
\Cam-8 is at heart a lattice gas machine.  Particle streaming is an
efficient, low-level hardware operation, and the large spatial data
shifts available make it convenient to investigate models with widely
varying particle speeds.  Multi-dimensional shifts are useful for
investigating models with shallow extra dimensions.

Our most advanced lattice gas collaboration is with Jeff Yepez and his
group at the U.S. Air Force's Phillips Laboratory.  He and Phillips
Labs have started a new initiative on geophysical simulation that
involves the construction of a large \cam-8 machine.

Geophysical phenomena are good candidates for lattice dynamics
modeling since there is so much distributed complexity involved, and
since many of these phenomena are so hard to model using traditional
differential equation techniques.  With lattice gases, the simulation
runs just as fast with the most complex boundary condition as with the
simplest.  One can use a great deal of physical intuition in
incorporating desired properties into models by constructing
simplified discrete versions of the actual physical dynamics.  The
process of making these models is closely akin to that of making
models in statistical mechanics, where one strives to include only the
essence of the phenomenon \cite{yepez-waterloo}.

Figures \ref{fig.fhp7} and \ref{fig.rolls} illustrate some simple
``warmup'' calculations done in collaboration with Yepez that
illustrate the use of \cam-8's statistics gathering hardware.  Here,
we split the system up into bins of a chosen size and use lookup
tables to count a function of the state of the sites in each bin.
These {\em event counts} are continuously reported back to the
workstation that is controlling the simulation.

Figure~\ref{fig.fhp7}a shows momentum flow in a two-dimensional
2K$\times$1K lattice, illustrating vortex shedding in lattice-gas flow
past a flat plate.  Here we use a 7-bit ``FHP'' model, which runs on
our prototype at a rate of 382 million site updates per second (for
pure simulation).  Both the time averages (over 100 steps) and the
space averages (over 32$\times$32 sites) were accumulated by \cam{};
the workstation simply drew the arrows.

Similarly, Figure~\ref{fig.fhp7}b uses the same model to illustrate a
Kelvin-Helmholtz shear instability on a 4K$\times$2K lattice.  Most of
the fluid was initially set in motion at Mach 0.4 to the right, except
for a narrow strip in the middle which was started with the opposite
velocity.  The Figure shows the situation after 40,000 time steps
(about 15 minutes of simulation).  The averaging is over regions of
128$\times$128 sites, and over 50 time steps.

\figgiven{rolls}{ \hfill%
\vbox{%
\psfig{figure=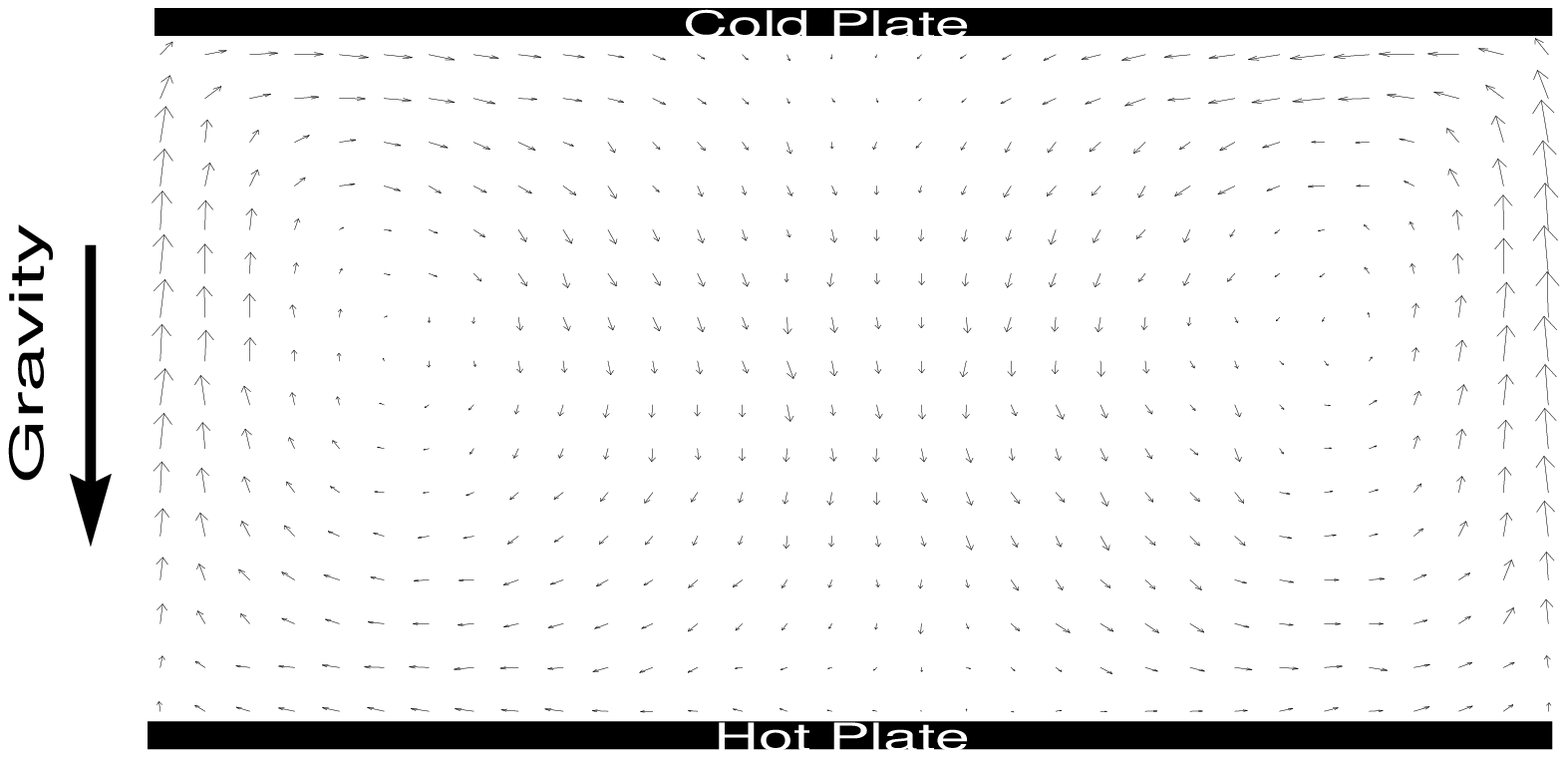,width=2.75in}%
}%
\hfill }{Rayleigh-B\'enard convection.}

Finally, Figure~\ref{fig.rolls} illustrates Rayleigh-B\'enard
convection, following \cite{chen-convection}.  The simulation uses a
13-bit hexagonal lattice-gas, with 3 particle speeds, heating (at the
bottom), cooling (at the top), walls around the box, and gravity.  The
simulation size is 1024$\times$512, and the prototype runs this at a
rate of 191 million site updates per second.  The time and space
averaging was done by \cam{} as in the previous figures.

We have also been working with Bruce Boghosian and Dan Rothman on
three dimensional lattice gas models.  Since the CM-2 also has 16-bit
lookup tables, the ``random isometry'' techniques that were used to
partition lattice-gas updates into a composition of 16-bit lookups
on the Connection Machine carry over directly to
\cam-8 \cite{dhumieres86a,boghosian-luts}: a 24-particle FCHC lattice
gas with solid boundaries runs at about 7 million site-updates per
second.  We are using these techniques as the basis for implementing
simulations of the flow of immiscible fluids through porous
media \cite{rothman92}.

\subsection{Statistical mechanics}

\figtwogiven{matter}{ \hfill%
\vbox{%
\psfig{figure=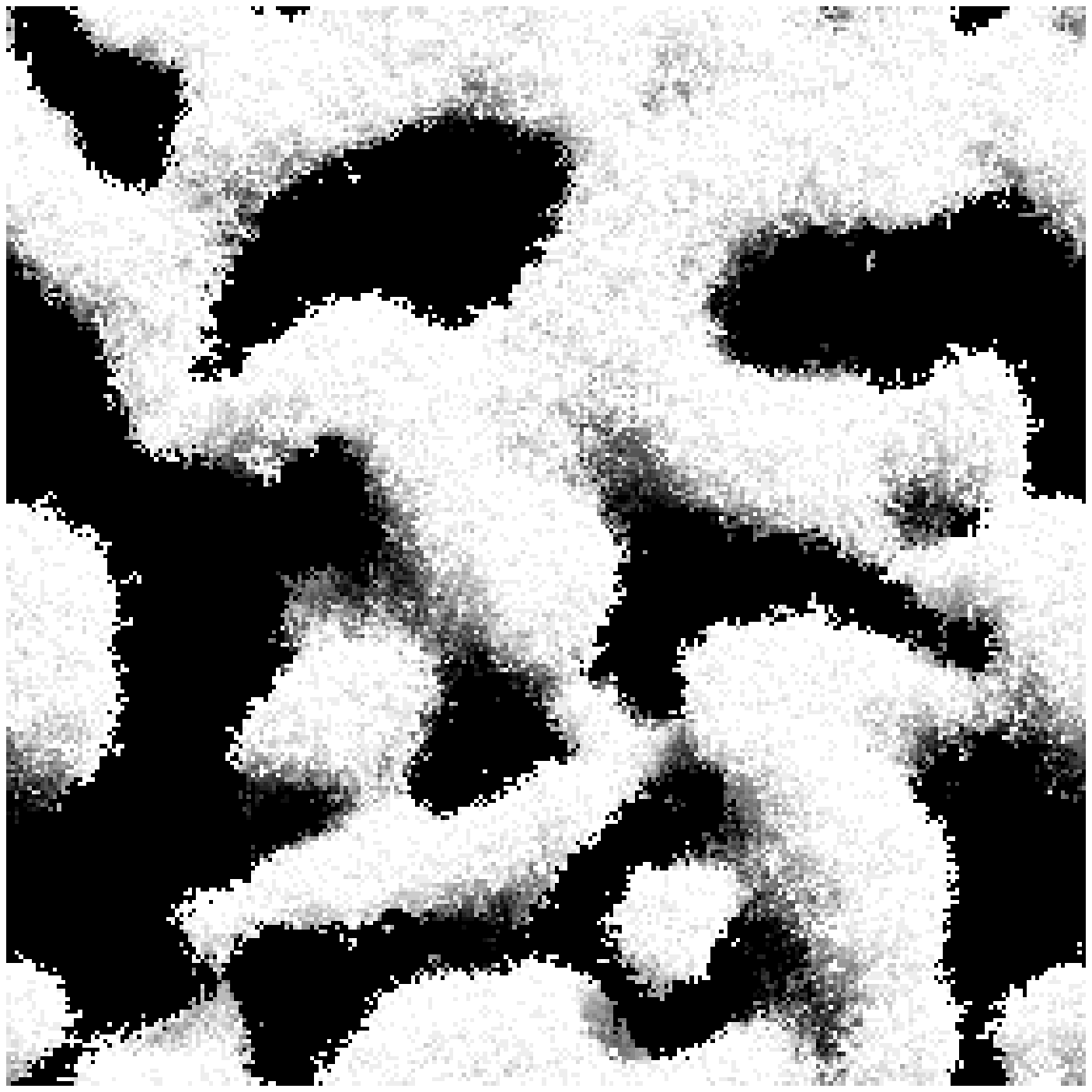,bbllx=99pt,bblly=200pt,bburx=499pt,%
bbury=599pt,height=2in}%
\vskip .3 in
\psfig{figure=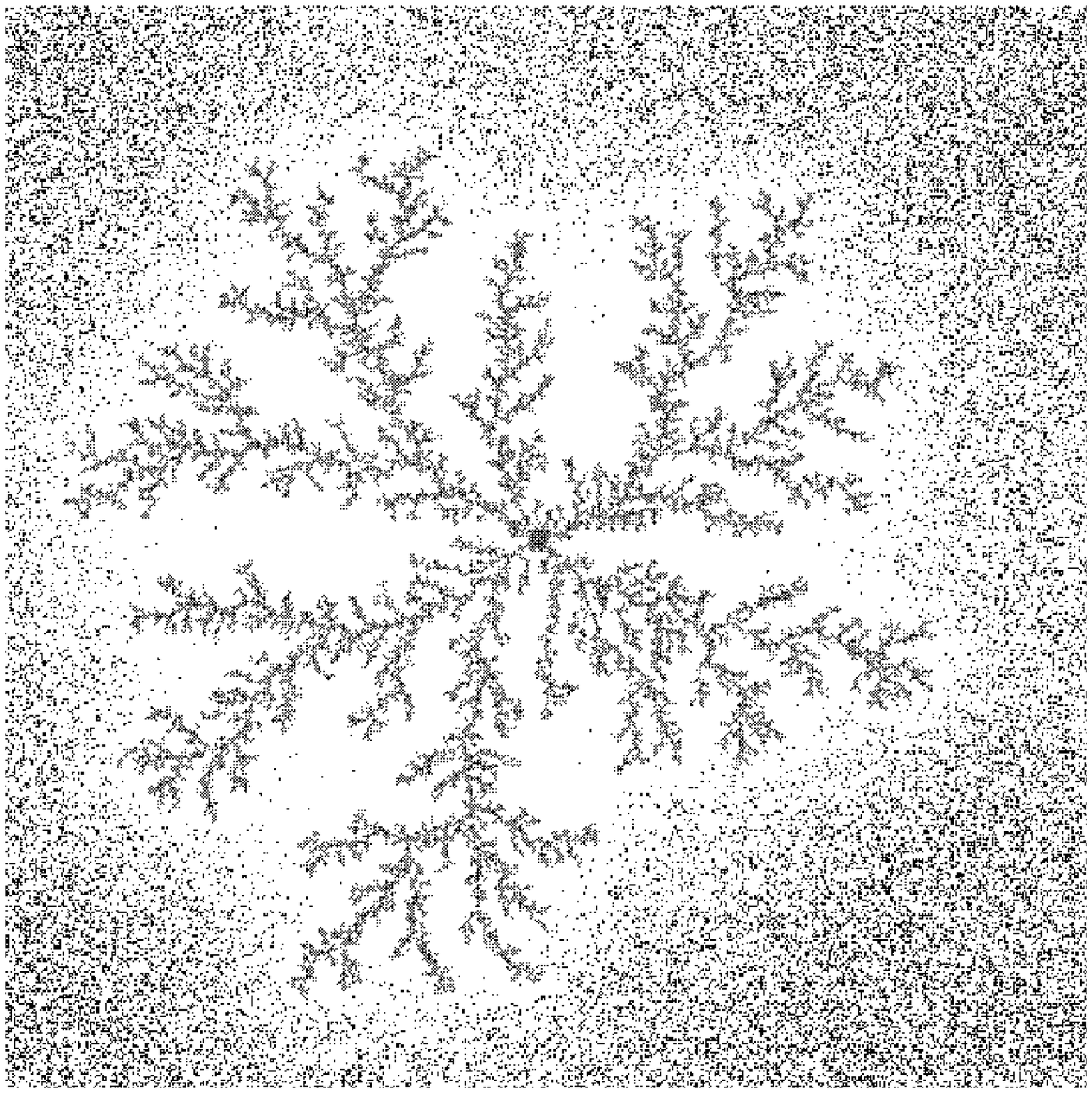,height=2in}%
}%
\hskip.3in
\vbox{%
\psfig{figure=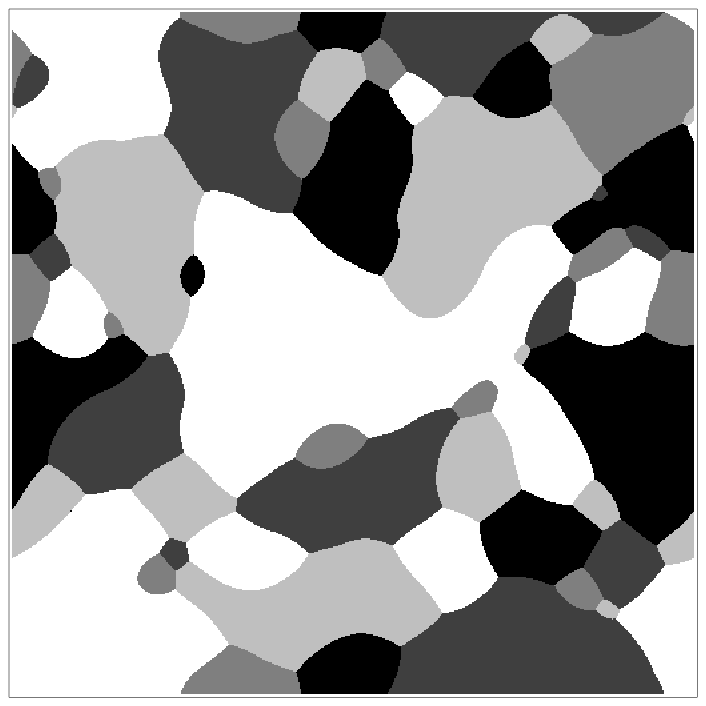,bbllx=72pt,bblly=521pt,bburx=270pt,%
bbury=720pt,height=2in}%
\vskip .3 in
\psfig{figure=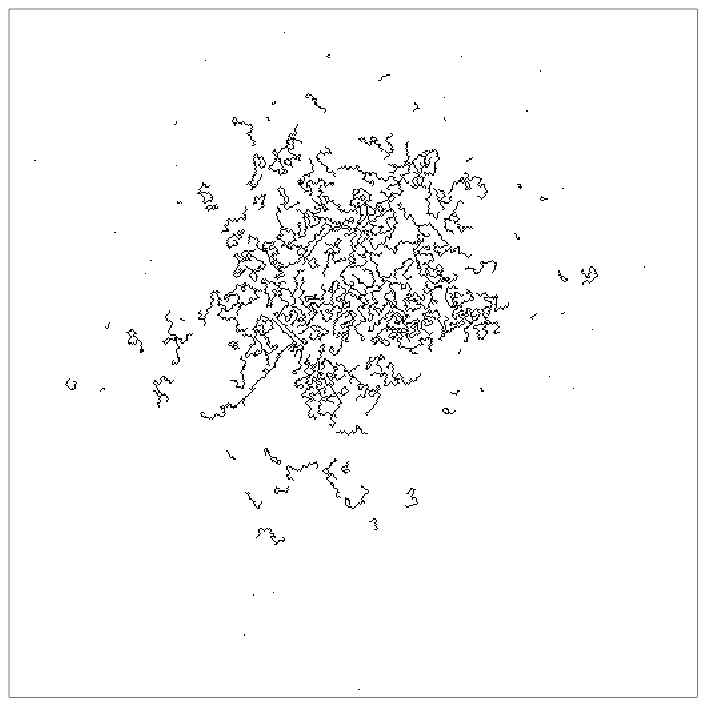,bbllx=72pt,bblly=521pt,bburx=270pt,%
bbury=720pt,height=2in}
}\hfill
}{Four materials simulations. (a) Spongy three-dimensional structure
obtained by ``majority'' annealing.  (b) Typical texture produced by
one of Griffeath's large-neighborhood voting rules. (c) Diffusion
limited aggregation.  (c) Polymers diffusing from an initial
concentrated region.}

\noindent
Physicists have long used discrete models in statistical mechanics to
model material systems.  In simulating such systems it is often
important to have available large quantities of precisely controllable
random variables.  On \cam-8, by independently applying large random
spatial shifts to each of a few randomly filled bit fields (and by
employing other related techniques), it is possible to avoid local
correlations and continuously generate high quality random variables
without slowing the simulation down.  Using such random variables, we
have run three dimensional thermalized annealing models
\cite{vichniac-anneal} on our 8-module prototype at about 200
million site-updates per second on a space of 16 million 16-bit sites
(about 12 updates of the 3-dimensional space per second), with
simultaneous rendering (by discrete ray tracing as part of the CA
dynamics) and display.  Figure~\ref{fig.matter}a shows one
rendered image from the \cam-8 display for a $512\times512\times64$
simulation.

Figure~\ref{fig.matter}b shows a deterministic simulation of a model
due to David Griffeath at the University of Wisconsin.  He and some of
his collaborators are engaged in the analysis of combinatorial
mathematics problems that have spatial locality.  They have been using
our earlier, much more limited \cam-6 machine in this capacity for a
number of years \cite{fisch}.  The simulation shown is a kind of
annealing rule: each site in the space ($512\times512$) takes on
whatever value is in the majority in its neighborhood.  The
neighborhoods are quite large---they involve the 121 neighbors in an
$11\times11$ region surrounding each site.  Since there are 5
different species (3 bits of state), the updating rule must deal with
363 bits of state in each neighborhood.  This is done as a composition
of about 70 distinct updating steps, and so we get about 10 complete
updates of the space per second (about 2.5 million site updates per
second).  A better algorithm, that doesn't recalculate the
species-counts for overlapping regions of adjacent neighborhoods,
would run an order of magnitude faster.  In either case, this example
serves to illustrate how rules that involve the interaction of large
numbers of bits at each site are handled by composing updating
steps.\footnote {At the opposite extreme of few bits per site, the
``Bonds Only'' \cite{cambook} version of Michael Creutz's dynamical
Ising model \cite{creutz} is a 1-bit per site partitioning rule that
runs at a rate of about 3 billion site updates per second on our
prototype.}

Figure~\ref{fig.matter}c shows a two-dimensional diffusion limited
aggregation simulation on a $1024\times1024$ space, driven by random
variables.  The system shown is started with a single fixed particle
in the center of about 100,000 randomly diffusing particles.  Whenever
a diffuser touches a fixed particle, it becomes fixed at that
position.  This is a large version of a \cam-6
experiment \cite{cambook}, but run more than two orders of magnitude
faster than \cam-6 could have run it.  The simulation performs about
800 million site updates per second, and the Figure shows the state of
the system after about two minutes of evolution.

Figure~\ref{fig.matter}d shows another statistical particle-based
simulation: a CA polymer model due to Yaneer Bar-Yam (
cf. \cite{yaneer-poly,smith92}; the \cam-8 program was written by
Michael Biafore).  This discrete model captures certain essential
features of polymers: conservation of the total number of monomers,
preservation of connectivity, monomers can't overlap (excluded
volume), etc.  It employs a statistical dynamics (controlled by \cam-8
random variables) that uses space-partitioning to maintain these
constraints \cite{margolus-bbm}.  The simulation discussed in
\cite{yaneer-poly} ran at a rate of about 30 million site updates per
second on a space $512\times512$.  Problems that are being addressed
with these models include dynamics in polymer melts, gelation and
phase separation, polymer collapse, and pulsed field gel
electrophoresis \cite{electrophoresis}.

\smallskip

\Cam-8 is designed to numerically analyze the models run on
it---largely through the use of the event counters \cite{cambook}.  By
appropriately augmenting the system dynamics with extra degrees of
freedom, we can make essentially any desired property of the system
quantitatively visible.  For example, localized spatial averages (such
as density, pressure, energy density, temperature, magnetization
density) can be gathered as we did to produce the momentum flows in
the previous section; global correlation statistics can be accumulated
quickly for occurrences of given spatial patterns; autocorrelations
can be computed by comparing the system to a copy of itself shifted in
time and/or space \cite{super}; and block-spin transformations can be
quickly performed, simplifying renormalization group calculations of
critical exponents.

\subsection{Data visualization and image processing}

Another area we've been exploring is two- and three-dimensional image
processing.  We were led into this area initially by the display needs
of our physical simulations (e.g., see Figure~\ref{fig.matter}a,
discussed above), but this activity has taken on a life of its own.

Our \cam-8 machine simulates a kind of raster-scan universe, in which
each hardware module sequentially scans its chunk of the overall
simulation space.  This raster scan can in fact be programmed to be
two dimensional, and synchronized and interfaced with an external
video source.  The necessary hardware is included as part of our
prototype, and allows us to perform realtime image processing.
Generic bit-map processing/smoothing/improving techniques are
supported through a combination of local (CA) operations and global
statistics gathering via the hardware event counters \cite{pratt}.
Well-known CA image-processing algorithms, such as those used
commercially for locating and counting objects in images, can also be
run efficiently \cite{rosenfeld,sternberg,wilson}.

\figtwogiven{rot}{ \hfill%
\vbox{%
\psfig{figure=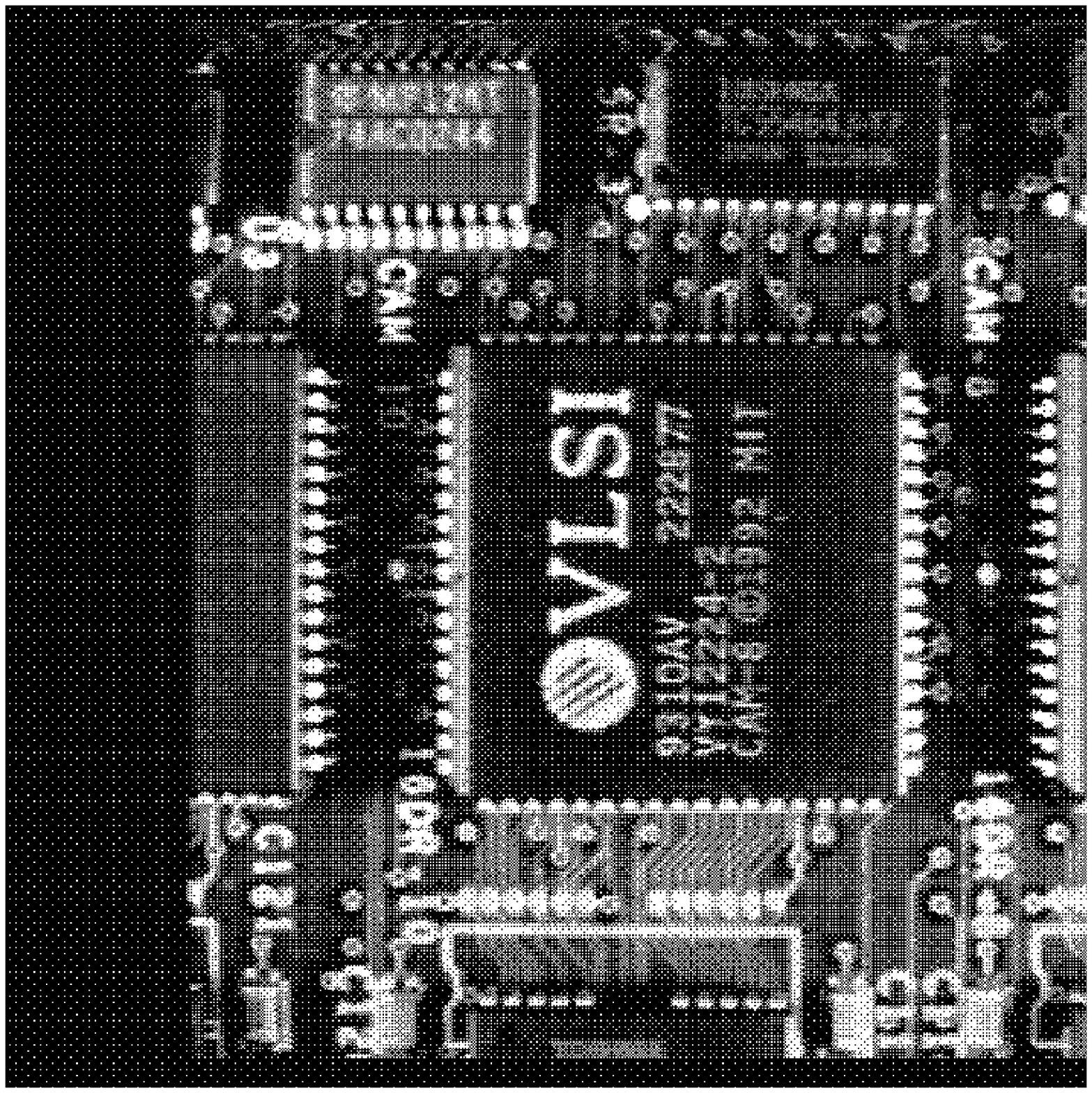,height=2in}
\vskip .3 in
\psfig{figure=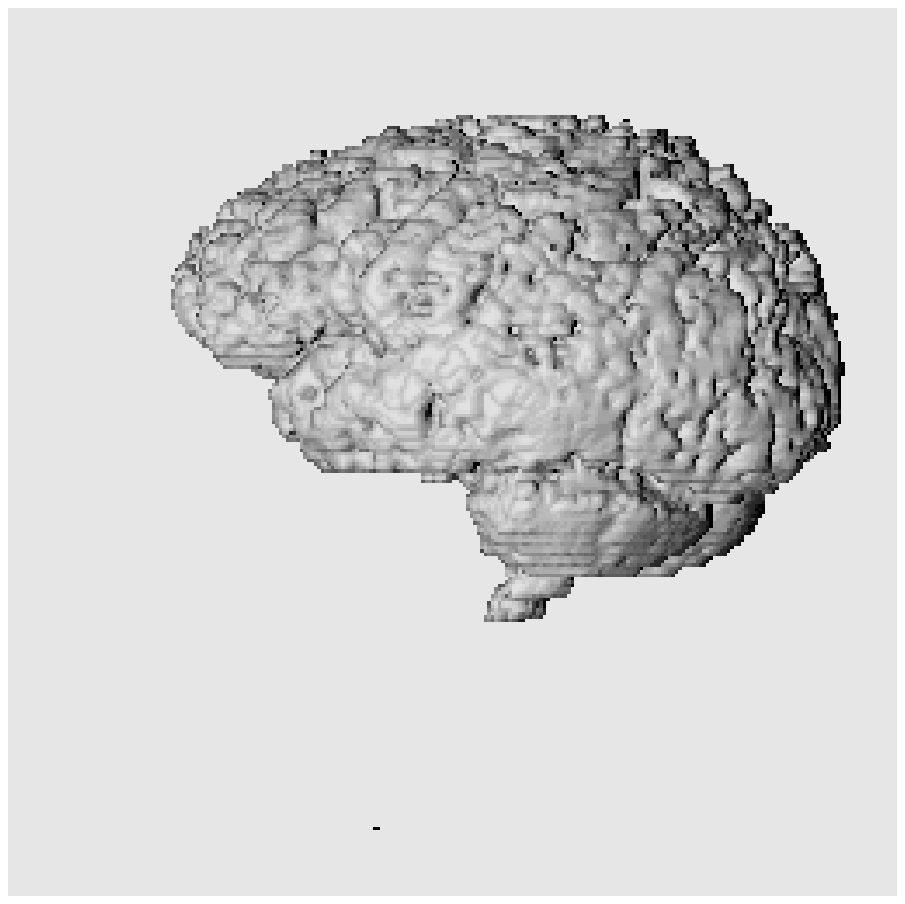,height=2in}
}%
\hskip .3 in
\vbox{%
\psfig{figure=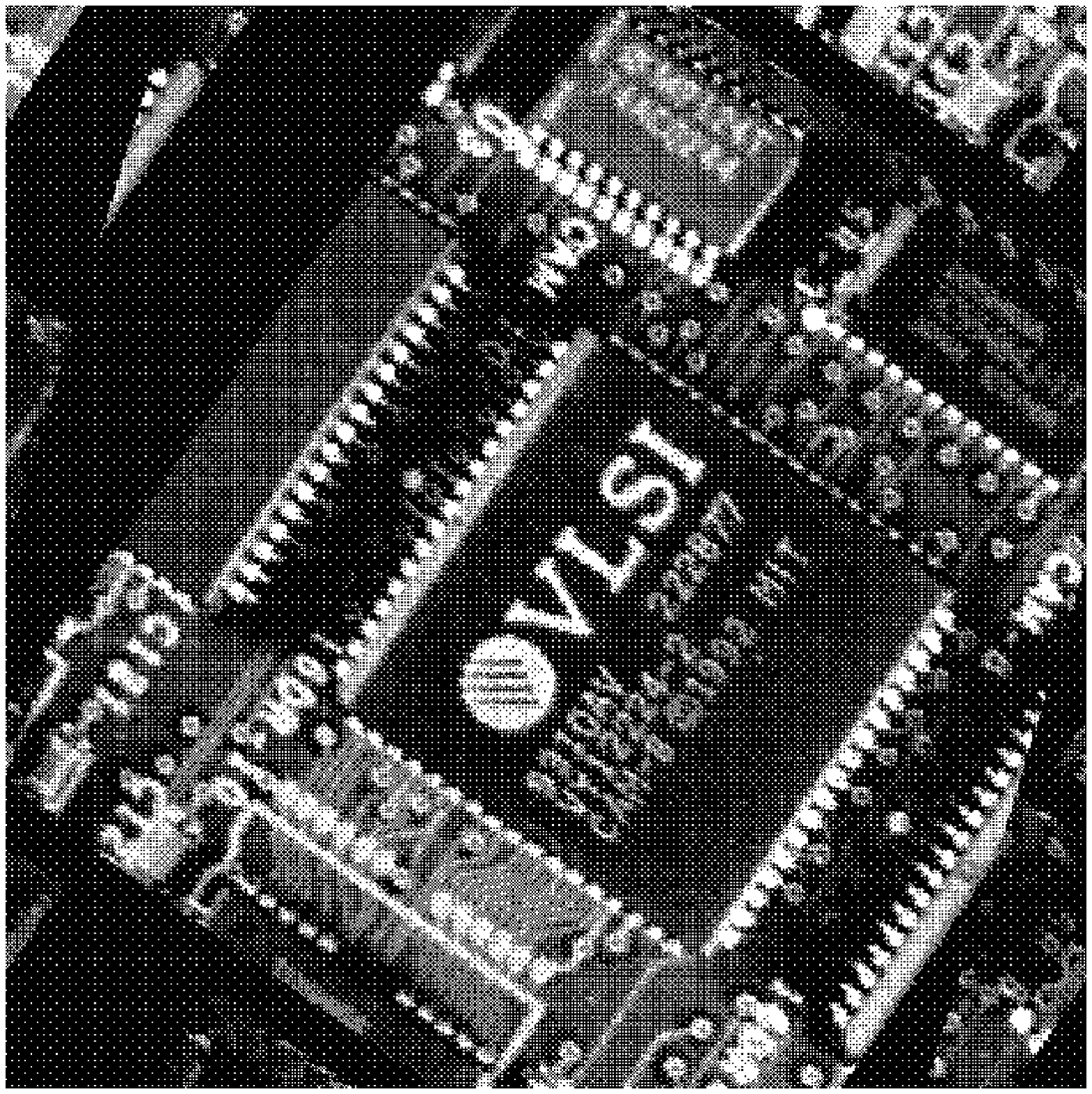,height=2in}%
\vskip .3 in
\psfig{figure=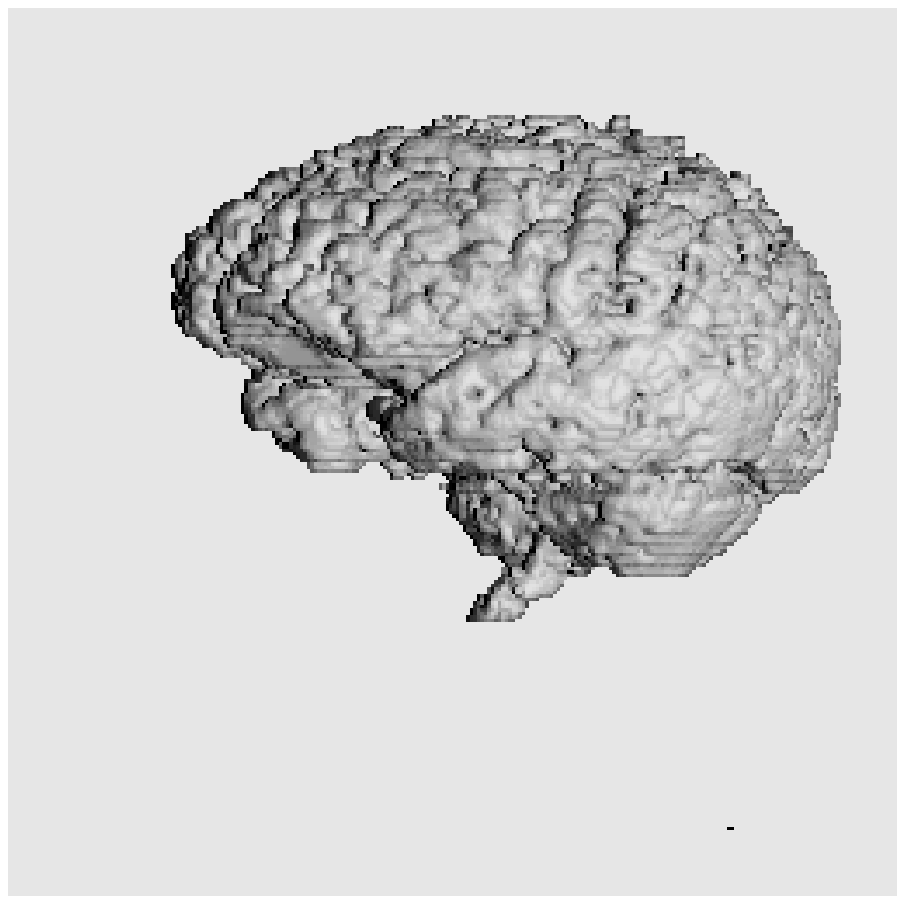,height=2in}%
}%
\hfill }%
{Continuous rotations on a CA machine.  Top: 2D rotation of realtime
video data.  Bottom: 3D rotation of MRI data.}

Many novel algorithms are also directly supported by the architecture.
For example, the 8-node prototype can rotate a $512\times512$ bit-map
image through an arbitrary angle in less than 10 milliseconds by
permuting the arrangement of the pixels to move every pixel to within
one pixel-width of its best possible rotated position.
Figure~\ref{fig.rot}a shows camera input of a closeup of the
\cam-8 chip (the semi-custom chip that knits memory chips together
into a CA machine).  Figure~\ref{fig.rot}b shows the same image
rotated by \cam-8 through an angle of 35
degrees \cite{graph-gems-rot}.\footnote{This same kind of technique is
applicable in other contexts.  For example, a matrix transpose can be
accomplished by a 90 degree rotation and a flip---this combined
operation on \cam-8 takes the same time as the rotation alone.  Some
of our collaborators (Bryant York and Leilei Bao at Northeastern
University) are performing combinatorial searches on \cam-8 by
applying these kinds of techniques to large multidimensional
matrices.}

In three dimensions, local CA techniques can be used to find and to
smooth two-dimensional surfaces to be visualized.  For example,
magnetic resonance imaging can produce three-dimensional arrays of
spatial density data that subsequently need to be visualized.
Interesting features might be the surface of the brain, the surfaces
of lesions, blood vessels, etc.  Local rules can be used to trace
features (e.g., blood vessels are regions connected to segments that
have already been identified as blood vessels) and to smooth surfaces
(e.g., using annealing rules that have surface tension).  Once a
surface has been distinguished, many bit-map oriented rendering
techniques are available.  The simplest is probably the same one used
in Figure~\ref{fig.matter}a: just simulate ``photons'' of light moving
{}from site to site, entering the system from one direction, and being
observed from another.  Figures~\ref{fig.rot}c and \ref{fig.rot}d show
the surface of the brain generated from MRI data, and rendered by such
a technique.  The two images are rotated versions of the same
data---we can actually do an arbitrary three-dimensional rotation of
site data using the same technique used in Figures~\ref{fig.rot}a and
\ref{fig.rot}b in just three updating scans of the space
\cite{tom-3drot}.

If you render a surface twice, once from each of two slightly
separated vantage points, you can quickly produce stereo pairs.  We
have tested this technique\footnote{Mike Biafore led this effort.} in
some of our physical simulations: we have run a version of the
three-dimensional annealing simulation pictured in
Figure~\ref{fig.matter}a while continuously generating such stereo
pairs, without slowing the simulation down at all.  Using this
technique to generate images from many vantage points, one can quickly
generate data needed for producing holograms from computer volumetric
data.\footnote{\Cam-8 should also be useful for reconstructing
three-dimensional surfaces from holographic data.  The algorithm
implemented by the {\sc horn} machine \cite{horn} should run faster on
our \cam-8 prototype than on the special-purpose {\sc horn} hardware.}

\subsection{Space\-time circuitry}

\Cam-8 can rapidly perform not only arbitrary rotations, but also
affine transformations on its data---the hardware can skip or
repeatedly scan sites during updating in order to rescale an image.
Actually, we can do far better than this: \cam-8 can perform {\em
arbitrary} rearrangements of bits, with any set of local, non-uniform
operations along the way.  To get an arbitrary transformation, you
simply simulate the right logic circuit!

\figtwogiven{logic}{ \hfill%
\vbox{%
\psfig{figure=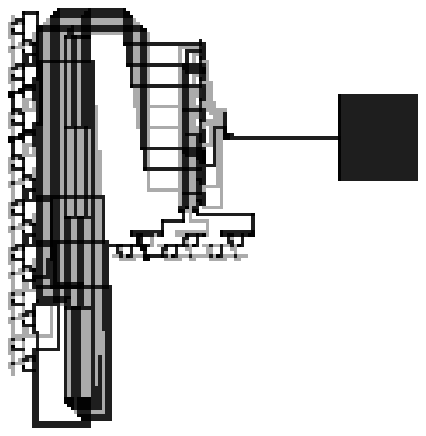,height=2in}
}%
\hskip .3 in
\vbox{%
\psfig{figure=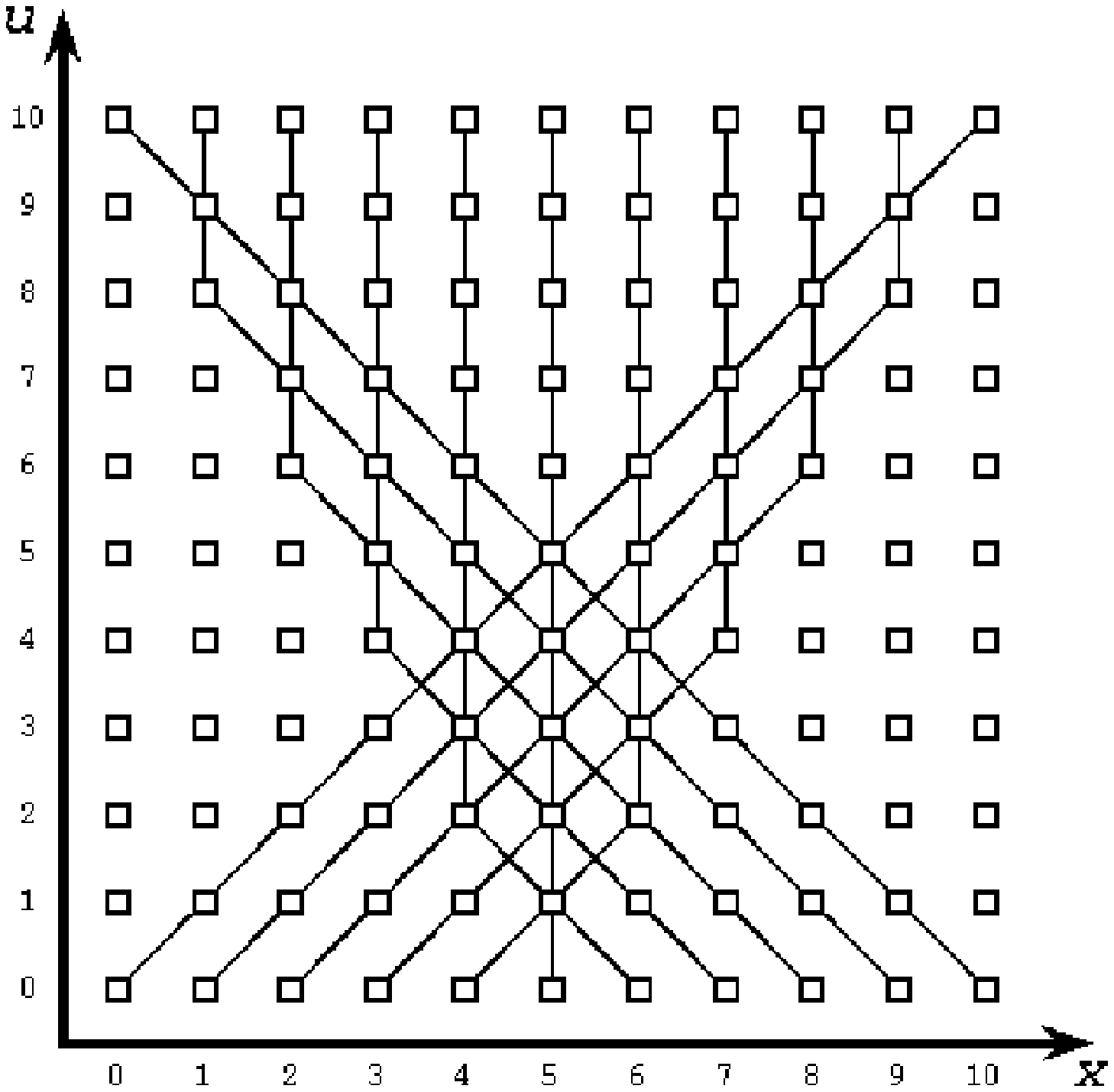,height=2in}%
}%
\hfill }{Logic simulation.  (a)~Gate-array-like CA simulation of a
random number generator.  (b)~Space\-time wires reverse the bits in a
1-dimensional space.}

A digital logic circuit is a physical system that (not surprisingly)
can be simulated efficiently by a (digital) CA space.
Figure~\ref{fig.logic}a shows a straightforward simulation of logic
using \cam-8.\footnote{The circuit shown is due to Ruben Agin.}  Here
we have a CA space that simulates a kind of sea-of-gates gate-array,
with one gate at each spatial site.  Local routing information
recorded at each site determines how data hops between bit-fields that
shift in various directions, in order to implement the wires that
connect the gates together.  Large three dimensional logic simulations
can be performed by \cam-8 in this manner: just as with other
spatially organized computations, the kind of virtualization of
spatial sites (gates here) that \cam-8 does makes such simulations
practical.\footnote{Since \cam-8 shares each processor over up to a
few million spatial sites, much higher performance specialized
machines with a lower virtualization ratio can be made to implement
specific \cam-8 rules---such as a logic simulation rule, or an image
processing rule.  You trade flexibility (large spatial shifts and
large lookup tables) and simulation size for speed.  Notice that even
if \fpga's are used for implementing these specialized machines, very
high silicon utilization ratios can be achieved, since the regular
structure of a CA maps naturally onto the regular structure of an
\fpga.} The investigation of CA rules that permit efficient logic
simulation is also important for highly-parallel fixed-rule CA
hardware: if the fixed rule supports logic simulation, then the
machine can simulate any other CA rule by tiling the simulation space
with appropriate blocks of logic.\footnote{The idea of using CA's to
do logic is quite old.  In fact, much of the present work on field
programmable gate arrays carries forward ideas that originated in
early work on CA's (cf. \cite{unger,hennie,minnick,shoup}.)}

\smallskip

Now consider the problem of producing rather general transformations
of the data in our CA space.  One approach would be to directly
simulate a gate-array-like rule that operates on the original data,
and eventually produces the transformed data.  An efficient technique
for doing this on \cam-8 is called {\em space\-time circuitry}.  This
involves adding an extra dimension to your system to hold the
transformation circuitry, laid out as a pipeline in which each stage
is evaluated only once, as the data passes through
it \cite{static,virt-wires}.

As a simple example, consider a 1-dimensional space where the desired
transformation is to reverse the order of the data bits across the
width of the space.  We add a dimension (labeled $u$ in
Figure~\ref{fig.logic}b) and draw a circuit that accomplishes the
reversal---in this simple case, we only need wires.  The circuit shown
is a data pipeline that copies information up one row at each stage,
and possibly over by one position right or left: the information about
which way the data should go is stored locally.  The \cam-8 rule that
achieves this transformation only involves 5-bit sites---two bits of
stationary {\em routing} information, and three shifting bit-fields to
transport the signals.  If we continually add new information at the
bottom of the picture, reversed data continually appears, with a
10-stage propagation delay, at the top.  But if we only want to
accomplish the transformation once, then we only need to update each
consecutive row of the circuit once, moving the signals up to the next
row before we update it in turn.  In this case, instead of one update
of the space moving the whole pipeline forward by one stage, the row
by row update will move one set of data all the way through the
pipeline!\footnote{The rendering algorithm of
Figure~\ref{fig.matter}a uses essentially this technique to
propagate the light all the way through the material system in a
single scan of the space.} We still get one result per update of the
space (exactly as before), but the propagation delay has been reduced
to a single scan of the space!  Thus given a CA space, by adding a
dimension containing a sufficiently complicated pipelined circuit, any
desired transformation of the original space can be achieved in one
scan of the augmented space---limited only by the total amount of
space available for the extra-dimensional circuitry.

If the problem we're interested in is the simulation of a clocked
logic circuit, this technique can be used to greatly speed up the
simulation.  Instead of updating our CA space over and over again
while signals propagate around the system, passing through gates and
eventually being latched in preparation for the next clock cycle, we
can pipeline this calculation using an extra dimension, and perform
the entire clock cycle in a single update of the space.  Since the
total volume of space (number of sites) needed to represent all of the
gates and wires should be comparable to the volume without the
pipeline dimension,\footnote{Since routing signals in a higher
dimension is generally much easier than in a lower dimension, the
circuit should actually be more compact.} this represents an enormous
speedup.  If we think of the routing and gate information that is
spread out in the pipeline dimension as being spread out in time, then
we greatly reduce the space needed for the calculation by making what
happens at each spot time dependent---hence the term {\em space\-time
circuitry}.

An additional benefit of space\-time circuitry on \cam-8 is that it
allows us to take good advantage of the large spatial shifts that are
available in this architecture.  In the logic example, we could use
big spatial shifts at some stages of the pipeline, and smaller ones at
other stages, in order to route all signals in as few stages as
possible---this provides a further speedup of the simulation.  Of
course these sorts of techniques (extra dimensions and big shifts)
will not be applicable to fully parallel CA machines built at the most
microscopic scale, but they add greatly to the power and flexibility
of our virtual-processor implementation.

\begin{figure*}[htbp]
  \centering
 {$
\vcenter{\hsize=2.2in\tiny\begin{verbatim}
new-experiment        512 by 512 space

                      0 0 == north  1 1 == south
                      2 2 == east   3 3 == west

define-rule hpp-rule  north south =  east west =  and
                      if east <-> north  west <-> south then
end-rule

define-step hpp-step  lut-data   hpp-rule
                      site-src   lut
                      lut-src    site
                      kick       north field -1 y
                                 south field  1 y
                                 east  field  1 x
                                 west  field -1 x
                      run        new-table
end-step\end{verbatim}}
\hskip 0.5in
\vcenter{\psfig{figure=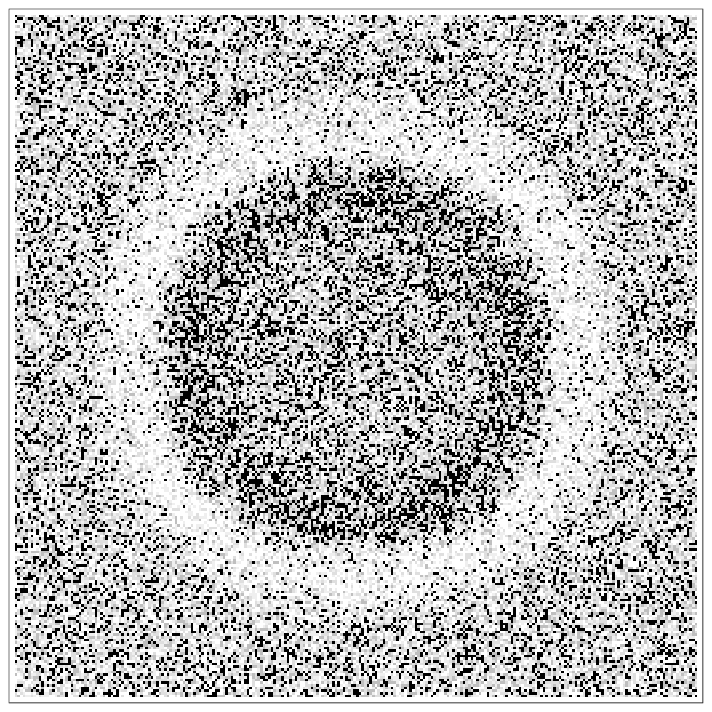,height=1.5in}}
 $}
\caption{Sample experiment.}
\label{fig.hpp-code}
\end{figure*}

\section{Software}

During the design of the \cam-8 \asic{}, we decided to implement a
version of the software that would drive the real hardware, and use
that to drive complete system simulations of the \cam-8 hardware,
including the workstation interface hardware.  Thus when the hardware
arrived, we immediately had software that would drive it, and could
run the same tests that we used to validate the design.

This initial software was intentionally rather low level, since it was
necessary to have low level access and control to thoroughly and
efficiently drive gate-level simulations that ran eight orders of
magnitude slower than a single module of the actual hardware.  The
present (still rather rudimentary) \cam-8 systems software has been
built as several layers on top of this initial work.  It provides a
prototypical programming environment for \cam-8 which demonstrates how
to access and control all facets of the hardware.

\subsection{A high level machine language}\label{high-level}

For simple CA models running on regular crystal lattices, the mapping
between the model and the \cam-8 architecture is quite
direct.\footnote{Embedding any regular lattice into \cam's Cartesian
lattice generally involves combining several adjacent sites of the
original lattice into one \cam{} site.}

To illustrate this direct mapping for the simplest lattice gas model,
Figure~\ref{fig.hpp-code} shows a \cam-8 assembly language program for
running the HPP lattice gas \cite{hpp}.  This program translates into
about a dozen \cam-8 machine-language instructions to be broadcast to
\cam.  It has two main parts: a rule definition, and a definition of
what constitutes an updating step.  The updating step broadcasts the
rule, adjusts some \cam{} data paths, specifies some uniform data
movements of the four bit-fields used to transport particles, and
initiates a scan of the space.  Despite being at such a low level,
this program runs without change on a machine with any number of
modules.\footnote {Utilities that download initial patterns and that
manage the video display are not shown here---the lowest-level
interaction of these routines with \cam{} depends explicitly on the
number of modules.} Issues such as making the data move smoothly
across module boundaries are handled directly by the hardware.

Figure~\ref{fig.hpp-code} also shows a ``snapshot'' from the \cam-8
display of a sound pulse resulting when this exact code is run from an
initial pattern of random particle data with a cavity (a 64$\times$64
particle vacuum) in the center.

\subsection{Zero-module scalability}

The \cam-8 machine language is directly interpreted by the hardware
interface that resides in the workstation that controls \cam.  This
language forms a sharp and simple boundary between the software and
the hardware---all interaction gets funneled through this interface.
A software simulator of \cam-8 has only to correctly interpret this
machine language in order to be compatible with all higher level
software written for \cam.

Since the \cam-8 architecture depends so heavily on data movement by
pointer manipulation and updating by table lookup, it is in fact very
well suited to direct software simulation on serial machines.  A
functionally accurate software simulator of \cam-8 has been
constructed for the Sun SPARCstation which runs CA models about as
fast as the best existing CA simulators for that machine---as fast as
simulators that are not burdened with the constraint of also
simulating \cam-8 functionality \cite{milan}.

This property that \cam-8 simulations have of running well even in a
pure software context we sometimes refer to as {\em zero-module
scalability}.  Efficient simulability on a variety of parallel and
serial architectures should encourage the use of the \cam-8 machine
model as a standard for CA work---which would make other \cam-8
software efforts much more widely useful.  Applications developed on
faithful software simulators (and on small \cam-8 installations) will
be directly transferable to large \cam-8 machines when faster or more
massive simulations are needed.

\subsection{Programming environment}

For specific applications, it will be the simulation context that
defines the ``high level'' programming environment.  For a logic
simulation, the high level environment might include hardware
description languages, logic synthesizers, chip-model libraries, etc.
For a fluid simulation, the high level environment might allow one to
``design'' a wind-tunnel, obstacles, probes, etc.  In general, one
needs facilities for conveniently producing interesting initial
conditions, for visualizing the state of the system, for monitoring
and analyzing the progress of the simulation, etc.  Our task here is
to provide examples, utilities, and ``hooks'' to facilitate the
construction and integration of such environments.

For developing models, one great simplification has been the sharing
of code that is possible between models that employ a similar spatial
format.  For example, we have constructed a set of libraries that
specialize the \cam-8 machine to run \cam-6 style neighborhoods on
variable-sized two-dimensional spaces.  This allows generic mechanisms
for display, analysis, etc. to be shared, allowing the programmer to
concentrate on developing models.  These libraries serve both to allow
the experiments and experience of \cam-6 to be applied rapidly to this
new domain and to allow users to develop applications in a simplified
and well documented context.  The library routines also serve as
examples of how to directly program \cam-8 itself.

The task of providing high-level tools for model development has
barely begun.  Some of the work involves only software engineering:
for example, writing good compilers that can automatically partition a
rule on sites with many bits into a composition of 16-bit operations
would be a valuable aid (cf. \cite{chortle}).  Compilers that can
perform specified transformations on a space by constructing space\-time
circuitry would be similarly valuable.  Access to arithmetic array
operations directly on \cam-8 would be useful not only for model
building, but for model analysis.  High level debugging tools that let
one quickly compare a model's behavior against expectations are
essential.  Where adequate models exist, work needs to be done on
parameterizing known modeling techniques and ways of combining models.

\subsection{Theoretical challenges}

Ideally, one would like to be able to specify a very high level
description of a physical system, and have software use some set of
correspondence rules to generate an efficient, fine-grained CA model
of that system.  In general, we don't know how to do this.  Present
modeling techniques are rather ad hoc, and the best progress has been
made by ``dressing up'' lattice gases by adding additional particle
species and interactions, resulting in complex models with large
numbers of bits at each site.  Such models are ill suited to an
ultimate goal of harnessing fine-grained, high-density microphysical
systems for CA computations.  Furthermore, there are at present no
fine-grained CA models of many basic physical phenomena, such as
motion of an elastic-solid, long-range forces, or relativistic
effects.

We know that more general methods of constructing models are possible.
For example, the numerical integration of a finite difference equation
is actually a type of CA computation, and it can reproduce a
differential equation.  This correspondence, however, yields a rather
restricted class of CA rules, constrained to use only arithmetic
operations and large numbers of bits at each site.  Without these
constraints, other general methods may be possible which yield much
simpler CA rules that also reproduce a desired macroscopic
dynamics---rules better suited to high-density microphysical
implementation.  Finding such general methods is an open problem.

Many basic questions remain in the development and analysis of CA
models, and progress on their resolution will both facilitate, and be
facilitated by, the use of CA machines.

\section{Conclusions}

By exploiting the uniformity of a virtual processor simulation of
fully parallel CA hardware, we were able to make workstation-class
hardware outperform supercomputers for many CA simulation tasks.
Using the same technology, a new generation of largescale CA machines
becomes possible that will make entirely new classes of spatially
organized computations practicable.  Our aim in all of this has been
to promote the development of CA models that can begin to harness the
astronomical computing power that is available, in a CA format, in
microphysics.

As stated, this goal is directed toward bringing computational models
closer to physics in order to improve computation, not physics.  But
computational models that match well with microphysics also tell us
something about the structure of information dynamics in physics.
Since a finite physical system has a finite entropy, not only computer
science but also physics itself must deal with the dynamics of
finite-information systems at increasingly microscopic
scales \cite{qc}.  Thus it seems possible that promoting the
development of physics-like computational models will one day
contribute to the conceptual development of physics itself.

\section*{Acknowledgments}

Many people contributed to the successful conclusion of the \cam-8
hardware project.  First I would like to thank our funding agency,
{\sc arpa}, which has strongly supported our CA machine research for
over a decade.  I would also like to acknowledge the intellectual debt
that this machine owes to Tom Toffoli, who built the first \cam{}
machine and had the basic idea of time-sharing a lookup-table
processor over a large array of cells.  He was also a close
collaborator in helping realize this design for a new CA machine,
along with Michael Biafore (intensive functional testing), Tom Cloney
(CM-2 simulator), Tom Durgavich (CAM chip design), Doug Faust and
David Harnanan (interface design), Nate Osgood (backplane), Milan Shah
(SPARC simulator), Mark Smith (early circuit prototyping), and Ken
Streeter (project programmer, and also prototyped part of our SBus
interface).  Harris Gilliam (project programmer), Frank Honor\'e and
Ken MacKenzie (board design), and David Zaig (technician) have been
helping get copies of our prototypes into shape for our collaborators.
In addition, I would like to thank Bill Dally and Tom Knight (MIT AI
Lab) for advice and design reviews; John Gage, Bruce Reichlen and Emil
Sarpa (Sun Microsystems) for help and encouragement; and Mike
Dertouzos and Al Vezza (MIT LCS) for their support.  Also, I'd like to
thank Jonathan Babb and Russ Tessier (LCS) for pointing out to me the
applicability of ideas about static routing to logic simulations on
\cam-8, and Gill Pratt, John Pezaris, and Steve Ward (LCS) for ideas
about how to actually build a large 3D mesh.  Finally, I'd like to
thank Mark Smith and Jeff Yepez for comments on this manuscript.

\bibliographystyle{}

\end{document}